# Cylindrical vector resonant modes achieved in planar photonic crystal cavities with enlarged air-holes

Kang Chang[1], Liang Fang[1], Chenyang Zhao[1], Jianlin Zhao[1,2], and Xuetao Gan[1,3]

[1]*MOE Key Laboratory of Material Physics and Chemistry under Extraordinary Conditions, and Shaanxi Key Laboratory of Optical Information Technology, School of Science, Northwestern Polytechnical University, Xi'an 710072, China*
[2]*jlzhao@nwpu.edu.cn*
[3]*xuetaogan@nwpu.edu.cn*

**Abstract:** We reveal a triangular-lattice planar photonic crystal supports Bloch modes with radially and azimuthally symmetric electric field distributions at the top band-edge of the first photonic band. Bifurcated from the corresponding Bloch modes, two cylindrical vector resonant modes are achieved by simply enlarging the central air-hole of the planar photonic crystal, which have high quality factors around 3,000 and small mode volume of $(\lambda/n)^3$. The far-field radiations of the two resonant modes present high-quality cylindrical vector beam profiles. The resonant modes could be optimized by modifying the six nearest neighboring air-holes around the central defect. The cylindrically symmetric characteristics of the resonant mode's near- and far-fields might provide a new view to investigate light-matter interactions and device developments in planar photonic crystal cavities.

## 1. Introduction

Planar photonic crystals (PPCs), characterized with periodic air-holes in a sub-wavelength thick slab, have great potentials for constructing chip-integrated functional devices and circuits because of the compact structure and engineerable photonic band structure [1]. When defects are introduced by missing, shifting, or shaping air-holes, the periodicity of PPC is broken and a PPC cavity is formed [2-3]. Determined by the high-efficiency Bragg reflection and sub-wavelength period of air-holes, resonant modes in PPC cavities exhibit ultrahigh quality ($Q$) factor and ultra-small mode volume ($V_{\mathrm{mode}}$), which have been widely employed in a variety of applications [4-6]. For instance, high-performance sensors [7], filters [8], and switches [9-10] were demonstrated by virtue of the shifted narrowband resonant peak. In nonlinear optical processes, PPC cavity's high $Q/V_{\mathrm{mode}}$ promises low-power continuous-wave pumped harmonic generations and wave mixings [11]. In addition, the ultrahigh density of photon state in PPC cavity is considered as one of the most effective routes to control classical and quantum light emissions, including low-threshold lasers and cavity quantum electrodynamics [12-14]. Note that these enhanced light-matter interactions strongly depend on the electric field directions of the resonant mode with respect to the nonlinear susceptibilities or emitter dipoles. For instance, in a quantum dot embedded PPC cavity, a single-hole missing defect was designed to optimize the alignment between cavity field and emitter dipole for achieving strong modulation of the spontaneous emission [15]. In a PPC cavity-enhanced second harmonic generation, crystal orientations of the gallium arsenide slab are considered to engineer the conversion efficiency and far-field patterns [16]. Therefore, electric field distributions of PPC cavity's resonant modes are crucial for their enhanced light-matter interactions and device developments.

Normally, PPC cavities with high $Q$ factors were designed by missing or shifting air-holes. Their resonant modes are standing-waves, showing electric fields with a $\pi$-phase variation between the neighboring intensity lobes. Thence, the directions of the electric fields over the intensity lobes are nearly parallel. A resonant mode with complex electric field directions is more desirable to expand PPC cavity's applications in strong light-matter interactions. Recently, cylindrical vector optical beams with spatially variant polarizations have attracted enormous research interests [17]. The unique high-numerical-aperture focusing properties of the cylindrical vector beams promise a variety of potential applications in nanoscale imaging [18], manipulations [19-20], and plasmon excitations [21].

Here, we report the achievements of two cylindrical vector resonant modes in PPC cavities, which have azimuthally and radially symmetric electric field distributions respectively. In a triangular-lattice PPC, we find Bloch modes with cylindrical vector characters are supported at the summit of the first photonic band. By simply enlarging the central air-hole, the corresponding defect modes could be bifurcated. While the $Q$ factors of these resonant modes are not high (~3,000) compared to those in other reported PPC cavities [4-6], their cylindrically symmetric characteristics may provide new aspects in enhanced light-matter interactions in PPC cavity to develop vector-mode based micro-lasers, nonlinear processes, and optical manipulations [22]. This could be considered as a unique feature of the proposed PPC cavity from other PPC cavities formed by missing or shifting air-holes. The far-field radiations of the two resonant modes show high-quality cylindrical vector beam profiles as well. In addition, this work may open an alternative way to generate cylindrical vector beams on a photonic chip, which has benefits in compact footprint and stability [23-24].

## 2. Cylindrical vector Bloch modes in planar photonic crystal

To be consistent with the mature fabrication technology of PPC devices, we study silicon-based PPCs with a refractive index of 3.5 at the telecom-band and a slab thickness of 220 nm. The triangular-lattice PPC has a lattice constant of $a$=400 nm and the radius of air hole is $r$=120 nm, as shown in Fig. 1(a). The photonic band structure and mode characters of the designed PPC are calculated via a three-dimensional finite element technique (COMSOL Multiphysics). In a PPC with air-holes in a dielectric slab, only optical modes with TE polarization direction, i.e. electric fields are parallel to the slab plane, exist photonic bandgap as well as resonant modes [1]. Hence, all of the simulations in this work are considered in the form of TE modes. Figure 1(b) displays the calculated band structure, where a forbidden band-gap is formed between the dielectric- and air-bands. The inset schematically shows the two-dimensional Brillouin zone with the definitions of Γ, K, and M points. By introducing defects in the PPC to break the index's periodicity, including missing, shifting air-holes, and modifying air-hole size, resonant modes could be localized with its frequency inside the band-gap. If the refractive index variation induced by the defect is negative (positive), resonant modes tend to bifurcate from the summit of the dielectric-band (valley of the air-band) with the field distributions similar to the corresponding Bloch modes.

The PPC's Bloch modes around the summit of the dielectric-band are calculated, locating at the high-symmetry point K with a frequency of $1.4972\times10^{14}$ Hz. A pair of degenerated Bloch modes with cylindrical vector characters are obtained, as displayed in Figs. 1(c) and 1(d), where the zoomed images are shown in the right column. The Bloch mode shown in Fig. 1(c) has an electric field distribution over the region between the central air-hole and the around six nearest neighboring air-holes. The black arrows indicate the direction of the electric field, showing an azimuthal symmetry with respect to the central hole. The electric field of another Bloch mode shown in Fig. 1(d) distributes around the central air-hole as well, but locates between the adjacent air-holes of the six nearest neighboring air-holes. The electric fields present a radially symmetry with respect to the central air-hole, as indicated by the black arrows.

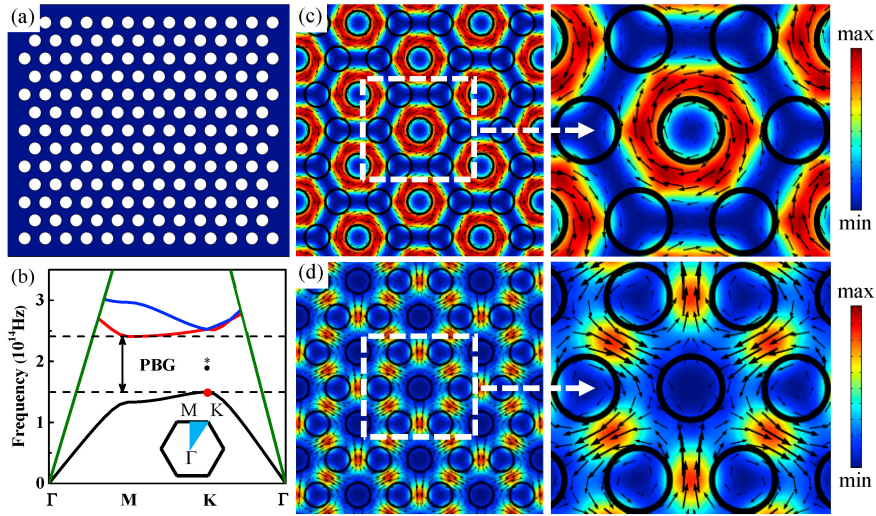

Fig. 1. Cylindrical vector Bloch modes in the PPC. (a) Schematic of the PPC structure with a triangular lattice of air-holes; (b) Photonic band structure of the PPC, where the red point represents the position of the two degenerated cylindrical vector Bloch modes and the green lines represent light cone; Inset displays the Brillouin zone with definitions of Γ, K, and M points; Black asterisk and black point indicate the frequencies of resonant modes in Fig. 2; (c-d) Electric field intensity of the (c) azimuthal and (d) radial Bloch modes with the zoomed images shown in the right column, where the black arrows indicate the electric field directions.

## 3. Cylindrical vector resonant modes in the planar photonic crystal cavity

The above demonstrated Bloch modes present an array of cylindrically symmetric vector modes, which may have potentials in applications employing on-chip large-scale vector electric fields. However, experimental excitation of these special Bloch modes is difficult and their confinements are weak. By introducing defects into the PPC to arouse resonant modes bifurcated from the two Bloch modes, it is possible to achieve a localized mode with similar electric field distributions but with stronger confinement.

In our design, to shift the frequency of the Bloch modes into the photonic band-gap and form a defect mode, we enlarge the central air-hole to induce a negative variation of the refractive index, as indicated in Fig. 2(a). The radius of the central air-hole is $R$. We calculate the resonant modes of the PPC cavity using the three-dimensional finite element technique. Only two resonant modes are obtained at the telecom-band in the designed PPC cavity and both of them have cylindrical symmetries. Figures 2(b) and 2(c) show the resonant modes of a PPC cavity with the central air-hole's radius of $R$=220 nm. The two modes are resonant at wavelengths of 1484 nm (Mode1) and 1591 nm (Mode2) with the $Q$ factors of 840 and 900, respectively, which are exactly located in the PBG region and represented by the black asterisk (Mode1) and black point (Mode2) in Fig. 1(b). The mode volumes are calculated as $0.234(\lambda/n)^3$ and $1.075(\lambda/n)^3$ for Mode1 and Mode2, respectively, where $\lambda$ denote the resonant wavelength in vacuum and $n$ is the refractive index of silicon slab. As shown in Fig. 2(b1), the electric field intensity of Mode1 has six peaks distributing symmetrically around the cavity center, and electric field directions represented by black arrows show azimuthal vector characters apparently. For Mode2, as shown in Fig. 2(c1), the electric fields also distribute symmetrically about the cavity center but are along radial directions at the angles of $\pi/6$, $\pi/2$, $5\pi/6$, $7\pi/6$, $3\pi/2$ and $11\pi/6$ in the polar coordinate. These two resonant modes have similar field distributions to the corresponding Bloch modes shown in Fig. 1.

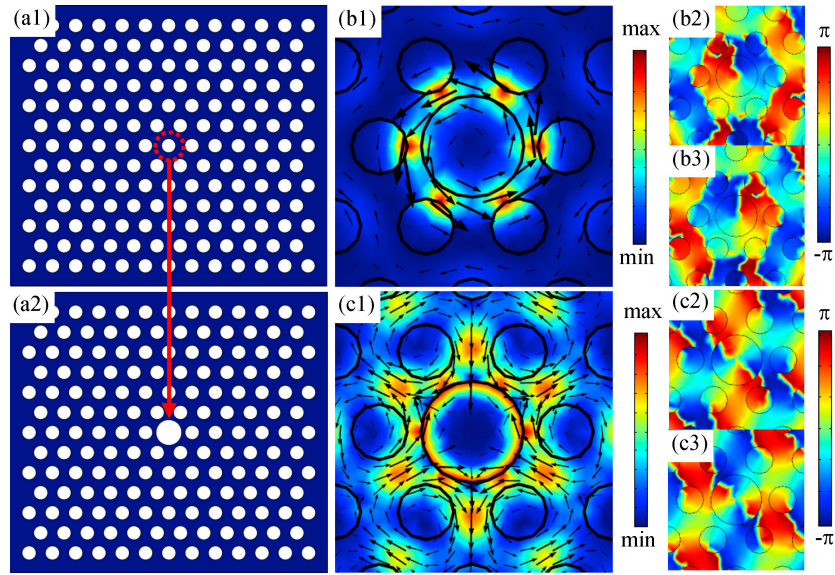

Fig. 2. Resonant modes in a PPC cavity with an enlarged central air-hole. (a1-a2) In-plane schematic of the PPC cavity; (b-c) Electric fields for (b) Mode1 and (c) Mode2, where (1)-(3) are electric field intensity, phase distributions of LH and RH circularly polarized components, respectively, and black arrows in (b1) and (c1) represent the electric filed directions.

To analyze the cylindrical vector characters of the two obtained resonant modes, we decompose their electric fields into left-hand (LH) and right-hand (RH) circularly polarized components using

$$E = E_x\hat{e}_x + E_y\hat{e}_y = (E_x - iE_y)(\hat{e}_x + i\hat{e}_y)/2 + (E_x + iE_y)(\hat{e}_x - i\hat{e}_y)/2 \qquad (1)$$

Where, $E_x$ and $E_y$ are the $x$ and $y$ components of the electric field in the Cartesian coordinate system, respectively; $\hat{e}_x$ and $\hat{e}_y$ are the corresponding orthogonal base vectors, and $(\hat{e}_x + i\hat{e}_y)/2$ and $(\hat{e}_x - i\hat{e}_y)/2$ represent LH and RH circularly polarized fields, respectively. According to the theory introduced in [25], both of the radial and azimuthal vector fields can be expressed by the superposition of LH and RH circularly polarized fields carrying vortex phase with opposite topological charges. Hence, it is crucial to examine the phase properties of the LH component ($E_x$-$iE_y$) and the RH component ($E_x$+$iE_y$).

For Mode1 shown in Fig. 2(b), the calculated results show that the amplitude difference between the LH and RH components can be ignored, and both of them have six maximum intensity points with approximately equal values. The phase diagrams of the LH and RH circularly polarized components are calculated, as shown in Figs. 2(b2) and 2(b3), respectively. For both of the two components, the phases have helical structures with a variation of $2\pi$ over the circumference around the central phase singularity. However, the screw directions of the helical phases are opposite for the LH and RH components. For sake of simplicity, we assume that the electric field amplitudes for both components at the six maximum points have a uniform value of $A$. And we can use $A\exp[-i(\varphi+\pi/2)]$ and $A\exp[i(\varphi+\pi/2)]$ to denote LH and RH components in the polar coordinate, respectively, where $\varphi$ represents azimuthal angle. According to superposition principle described in Eq. (1), we can obtain the electric field expression of Mode1 at the six maximum points as

$$E = A\exp[-i(\varphi+\pi/2)](\hat{e}_x + i\hat{e}_y)/2 + A\exp[i(\varphi+\pi/2)](\hat{e}_x - i\hat{e}_y)/2 = A(-\sin\varphi\,\hat{e}_x + \cos\varphi\,\hat{e}_y) \quad (2)$$

This is exactly an expression of standard azimuthal vector fields.

For Mode2, the phase distributions of LH and RH circularly polarized components are calculated and shown in Figs. 2(c2) and (c3). Helical phase structures are obtained as well for both components with opposite screw directions. Different from the phase structures obtained in Mode1, the vortex phase of Mode2 has a phase variation of $4\pi$ over the whole circumference, indicating a topological charge of 2 [24]. In a similar way, we can obtain the expression of the electric field as

$$E = A\exp[i(2\varphi - 2\pi/3)](\hat{e}_x + i\hat{e}_y)/2 + A\exp[-i(2\varphi - \pi/3)](\hat{e}_x - i\hat{e}_y)/2 \quad (3)$$

where $A$ denotes the amplitudes of electric fields at angles of $\pi/6$, $\pi/2$, $5\pi/6$, $7\pi/6$, $3\pi/2$ and $11\pi/6$ in the polar coordinate, and $\varphi$ denotes azimuthal angle. It is easy to demonstrate that the electric field directions are exactly along radial directions at the six angles, which is consistent with Fig. 2(c1). Considering that the topological charges of LH and RH components are -1 and 1 for Mode1, and 2 and -2 for Mode2, respectively, we define Mode1 and Mode2 as the first- and second-order cylindrical vector mode.

The cylindrical vector resonant modes in PPC cavities could be employed as a compact on-chip vector beam generator via their far-field radiations [23, 26]. By using Rayleigh-Sommerfeld diffraction integral theory [26], we calculate the far-field radiation patterns of Mode1 and Mode2. As show in Fig. 3(a1), the inhomogeneous electric field intensity of Mode1 evolves into a homogeneous doughnut distribution with a perfect azimuthal alignment of the electric field around the beam center. And the far-field radiation of Mode2 shows nearly doughnut shape with six lobes, as shown in Fig. 3(b1). The electric fields on the lobes are along the radial directions. The diffraction fields of the two resonant modes both show more purely cylindrical vector characters than the near fields. The phase diagrams in the LH and RH circularly polarized components of the far-fields are shown in Figs. 3(a2)-3(a3) and 3(b2)-3(b3). The topological charges of circularly polarized components maintain unchanged during the diffraction process. Figures 3(c) and 3(d) display polarization states of the far-fields shown in Figs. 3(a1) and 3(b1), respectively. The major axis' orientations of polarization ellipse represented by short lines, which further illustrate the azimuthal vector character of Mode1 and radial character of Mode2. The zero background illustrates the locally linear polarization characters of the radiation field.

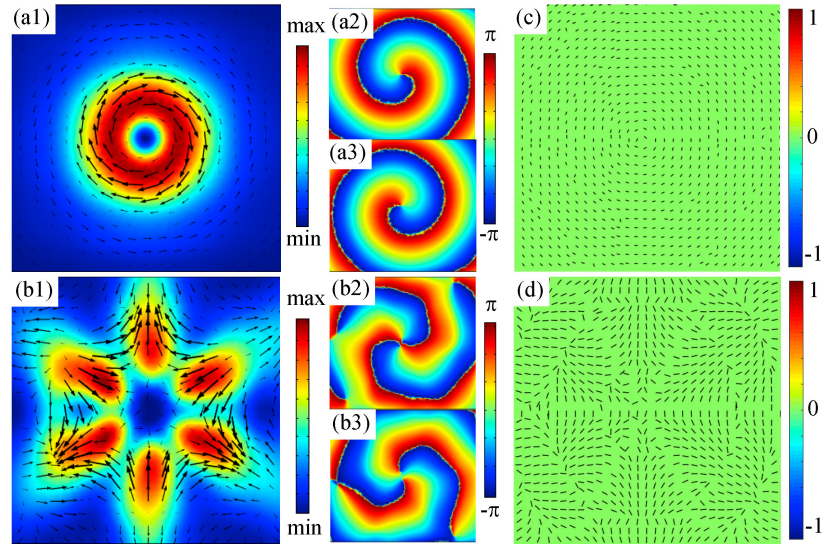

Fig. 3. Far-field radiation patterns of the cylindrical vector resonant modes. (a-b) Electric field for (a) Mode1 and (b) Mode2, where (1)-(3) are electric field intensity, LH and RH circularly polarized components' phase distributions, respectively, and black arrows in (a1) and (b1) represent the electric filed directions; (c-d) Polarization states of (c) Mode1 and (d) Mode2 in far-field, where the short lines and background denote orientation of major axis and ellipticity of polarization ellipse, respectively.

Since the proposed PPC cavity is formed by enlarging the central air-hole, the resonant modes depend on the central air-hole radius $R$ strongly. By varying the radius $R$, we simulate the resonant modes of the PPC cavity. The simulation results show that two cylindrical vector resonant modes discussed above exist for various $R$ between 130 nm and 250 nm, and no other resonant mode present at the telecom-band. The $Q$ factors, resonant wavelengths and mode volumes of the two modes are effectively tuned by $R$. As shown in Fig. 4(a), as the radius $R$ is increased from 140 nm to 250 nm, the first-order mode attains a maximum $Q$ value of 2,000 at $R$=193 nm, while the second-order mode achieves a maximum $Q$ value of 2,300 at $R$=245 nm. Different from $Q$ factors, the resonant wavelengths of the two modes both decrease monotonously in the process of increasing $R$, as show in Fig. 4(b). For the PPC cavity with larger $R$, the resonant wavelengths of the two resonant modes have larger difference. We can therefore selectively excite the desirable specific mode by tuning the laser wavelength. In addition, as shown in Fig. 4(c), the mode volumes of the two resonant modes also decrease monotonously with increased $R$, and attain minimum values of 0.2 $(\lambda/n)^3$ and 0.47 $(\lambda/n)^3$ for the first- and second-order modes at $R$=250 nm, respectively. Note, we also calculate the resonant modes of the PPC cavity with shrunk central air-hole ($R$<120 nm), no any cylindrical vector resonant modes are obtained.

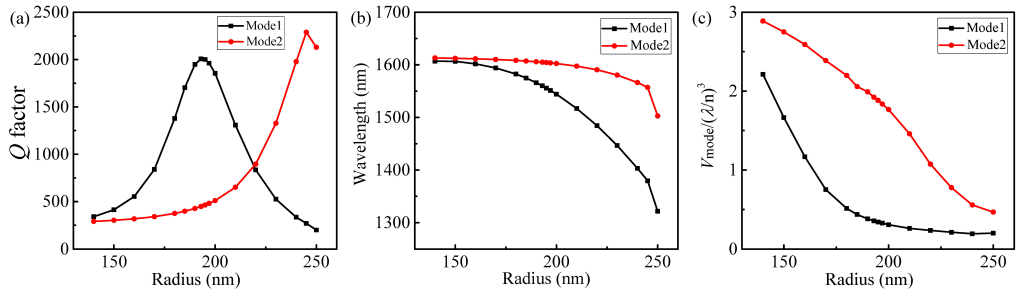

Fig. 4. (a) $Q$ factors, (b) resonant wavelengths and (c) mode volumes of the two resonant modes with respect to the radius ($R$) of the central air-hole.

## 4. Optimizations of the cylindrical vector resonant modes

While cylindrical vector resonant modes are bifurcated from the Bloch modes by enlarging the central air-hole, the cylindrical symmetries of the resonant mode are not as good as those of the Bloch modes, especially for the Mode2. To improve that, we enlarge the six nearest neighboring air-holes around the central defect to have the radius of $R$ as well, as displayed in Fig. 5(a). In the modified cavity structure, the defect region has seven enlarged air-holes with the same radius, which resembles to the homogenous PPC lattice shown in Fig. 1(a). The calculations of the resonant modes in the modified PPC cavity indicate that there only exist two resonant modes with azimuthally and radially symmetric field distributions. These two modes also bifurcate from the Bloch modes shown in Fig. 1, since the Bloch modes are only determined by the PPC lattice and material's dielectric constant and would not be changed by the cavity defect. For $R$ =150 nm, the calculated two resonant modes are shown in Figs. 5(b) and 5(c), which have resonant wavelengths of 1559.7 nm and 1558.9 nm, respectively. The mode volumes of the two modes are calculated as well with values of 0.33 $(\lambda/n)^3$ and 0.69 $(\lambda/n)^3$. For the azimuthal vector resonant mode in Fig. 5(b), the field presents a more homogenous doughnut distribution than that shown in Fig. 2(b1). In Fig. 5(c), strong electric fields are mainly concentrated between the adjacent air-holes of the surrounding six air-holes, which is similar to the Bloch mode shown in Fig. 1(d).

The modification of the PPC cavity not only optimizes the mode distributions to be more cylindrically symmetric, but also improves the $Q$ factors and decreases mode volumes of the two resonant modes simultaneously. Figure 5(d) plots the calculated $Q$ factors and mode volumes with respect to the radius $R$ of the seven enlarged air-holes. It shows that the two resonant modes achieve their maximum $Q$ factors when $R$ is around 150 nm. And the maximum $Q$ factors are 2,900 and 3,500 for the azimuthal and radial vector resonant modes, respectively. It is different from the resonant modes in the PPC cavity only with the central air-hole enlarged, which can't attain comparable large $Q$ factors simultaneously. These improvements could be explained as follows. In Figs. 1(c) and 1(d), a unit cell of the cylindrical vector Bloch mode comprises seven air-holes with same radius. It has been demonstrated the desirable vector resonant modes are bifurcated from these Bloch modes. For the PPC cavities with one or seven enlarged air-holes, the resonant modes are both distributed around the central seven air-holes. The PPC cavity with seven enlarged air-holes makes the central defect region with seven homogenous air-holes, which is similar to the PPC lattice shown in Fig. 1(a). Hence, the refractive index variation around the location of the resonant mode is much gentler, reducing the light scattering caused by the mismatch of the dielectric constant and therefore increasing the *Q* factor. This is one of the most fundamental concepts to raise *Q* factors in PPC cavity designs, as demonstrated in [2-5]. In addition, as shown in Fig. 5(d), the mode volumes of two resonant modes both decrease monotonously with the increased *R*. The mode volumes attain minimum values of 0.17 $(\lambda/n)^3$ and 0.34 $(\lambda/n)^3$ for Mode1 and Mode2 at *R*=180 nm, respectively, which are smaller than those obtained in the cavity shown in Fig. 2. This decreased mode volumes could be attributed to the strong and uniform concentration of light between the air-holes in the homogenous seven enlarged air-holes. Figures 5(e) and 5(f) display the far-field radiation patterns of the two resonant modes. In virtue of the improved cylindrically symmetric mode distribution, the far-field beam profiles have better symmetries as well.

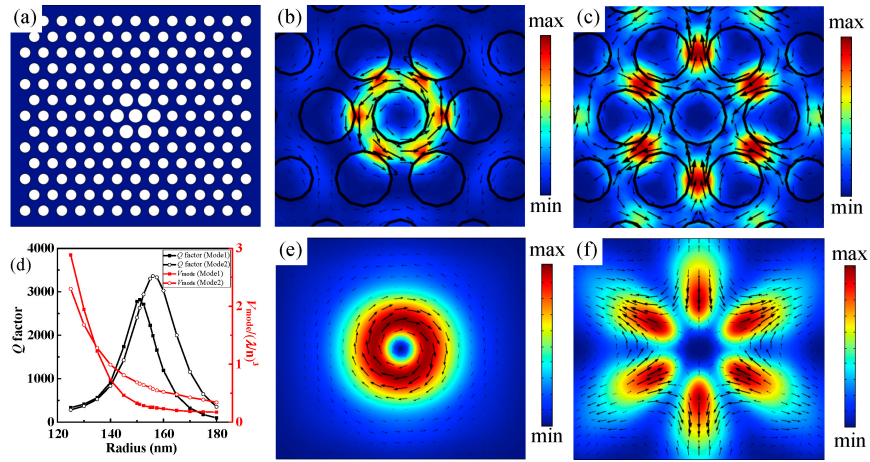

Fig. 5. Resonant modes in modified PPC cavity with seven enlarged air-holes. (a) In-plane schematic of the PPC cavity with seven enlarged air-holes; (b-c) Electric field intensity distributions of the (b) azimuthal and (c) radial vector resonant modes; (d) *Q* factors and mode volumes of the two resonant modes with respect to the radius of the seven enlarged air-holes; (e-f) Far-field radiation patterns for the two resonant modes, where black arrows represent electric filed directions.

## 5. Conclusion

In summary, we realize the generation of cylindrical vector resonant modes with high-*Q* factors and small mode volumes in PPC cavities formed by enlarging the central air-hole. The resonant modes are bifurcated from the Bloch modes at the summit of the photonic dielectric-band. Analyzing from the directions of electric fields and phase structures of the LH and RH circularly polarized components, we observe that the two modes have azimuthally and radially symmetric distribution. In addition, by enlarging the six nearest neighboring air-holes of the PPC cavity, *Q* factors and mode distributions of the cylindrical vector resonant modes are improved. The far-field radiations of the resonant modes are calculated as well, showing high quality cylindrical vector beam profiles. The cylindrical vector resonant modes generated in the PPC cavities and their far-field beams provide great potentials for chip-integrated generator of vector beams and enhanced light-matter interactions with complex electric fields.

## 7. Funding

This work was supported by the National Natural Science Foundations of China (11634010, 61522507, 61377035, and 11404264), the Key Research and Development Program (2017YFA0303800), and the Natural Science Basic Research Plan in Shaanxi Province of China (2016JQ6004).